\documentclass[]{aastex631}
\usepackage{xcolor}
\usepackage{longtable}

\newcommand{\pdv}[2]{\frac{\partial{#1}}{\partial{#2}}}

\newcommand{\revOne}[2]{#2}

\begin{document}

\title{Neutrino detection rates from lepto-hadronic model simulations of bright blazar flares}

\author[0000-0001-6399-3001]{Joshua Robinson}
\affiliation{Centre for Space Research, North-West University, Potchefstroom, 2520, South Africa}

\author[0000-0002-8434-5692]{Markus B\"ottcher}
\affiliation{Centre for Space Research, North-West University, Potchefstroom, 2520, South Africa}



\begin{abstract}

There is mounting evidence that blazars are the sources of part of the very-high-energy astrophysical neutrino flux detected by IceCube. In particular, there have been several spatial and temporal coincidences of individual IceCube neutrino events with flaring blazars, the most prominent of them being IceCube-170922A, coincident with a multi-wavelength flare of TXS~0506+056. Motivated by this, we used the time-dependent lepto-hadronic code OneHaLe to model the spectral energy distributions and light curves of a sample of bright $\gamma$-ray flares of blazars detected by Fermi-LAT, for which Kreter et al. (2020) provided calorimetric estimates of the expected neutrino detection rates. Flares were modelled with temporal changes of the proton injection spectra. Our analysis shows that the calorimetric approach overestimates the increase in neutrino production by a factor of typically $\sim 10$ if the $\gamma$-ray emission is dominated by proton-synchrotron radiation. 
\end{abstract}

\keywords{}


\section{Introduction} \label{sec:intro}
The sources of very-high-energy and ultra-high-energy cosmic rays remain unclear, with several plausible candidate source classes suggested that fulfil the Hillas criterion \citep{Hillas1984}.
The sites where cosmic rays are accelerated would also produce neutrinos due to the interaction between cosmic rays and either radiation fields or matter.
These neutrinos may be used to locate where the acceleration is taking place as the weak interaction cross sections coupled with the charge neutrality of neutrinos allow them to travel through both matter and magnetic fields without absorption or deflection \citep{Gaisser1994}. 
The jets of active galactic nuclei (AGNs) offer a suitable site for the acceleration of such high-energy neutrinos \citep[e.g.,][]{Mannheim1992} as the high luminosity within an emission region of relatively compact size provides a dense target photon field for photo-pion interactions and hence a suitable environment for photo-pion-induced neutrino production.
Analyses have, however, shown that \revOne{4}{gamma-ray bright blazars} are unlikely to be the dominant source of the diffuse neutrino flux detected by IceCube
 \citep[e.g.,][]{Murase2018}. Work by \cite{Yoshida2023} also shows that, even with flaring states taken into account, blazars cannot account for the bulk of the IceCube neutrino flux. 
\revOne{5}{Evidence of a spatial correlation between neutrinos and $\gamma$-ray loud \citep[e.g.,][]{Garrappa2019, Buson2022} blazars, as well as radio-loud, core-dominated blazars \citep[e.g.,][]{Plavin2020, Plavin2021, Plavin2023} is mounting}, and there have been numerous tentative associations of individual IceCube neutrino events with multi-wavelength flaring blazars. Most prominently, in 2017, IceCube detected a 290 TeV neutrino from the direction of the blazar TXS 0506+056 \citep{Aartsen2018}. At the time of the detection, TXS 0506+056 was in a state of increased $\gamma$-ray emission. Another remarkable example was the blazar PKS 0735+17, which was found in coincidence with not only the IceCube event IceCube-211208A, but potentially also neutrino detections by 3 other experiments (Baikal-GVD, Baksan, KM3NeT) within about a week of the IceCube alert \citep{Sahakyan2023, Acharyya2023}.
The active galaxy NGC 1068 was also identified as a source of astrophysical neutrinos detected by IceCube \citep{Abbasi2022}, further strengthening the link between AGN and neutrino production, though NGC 1068 is not a blazar, but rather a Seyfert II galaxy \citep{Seyfert1943}.

In \cite{Kreter2020}, neutrino detection rate expectations from a sample of the \revOne{1}{50 brightest $\gamma$-ray flares of blazars (at the time)}
were calculated using a calorimetric approach developed in \cite{Krauss2014} and \cite{Kadler2016}. 
In this work, we analyse the same flares as \cite{Kreter2020} using detailed physical simulations performed with the OneHaLe code \citep{Diltz2016, Zacharias2021, Zacharias2022}.
Archival multiwavelength data was used to construct spectral energy distributions (SEDs) of the blazars in their quiescent state. These were modelled with OneHaLe in a time-independent equilibrium state. 
Fermi-LAT data was used to construct both light curves and SEDs of these blazars in flaring states. Where available, Swift/XRT data was also used to provide additional contemporaneous SED and light curve information.
The flares were reproduced by perturbing the input parameters governing the proton injection spectrum and luminosity from the quiescent-state fits on a timescale similar to that of the length of the flares to simultaneously reproduce the flaring-state SEDs and the Fermi-LAT $\gamma$-ray light curves.

The structure of this paper is as follows. In Section \ref{sec:data} we outline the methods used to analyse each flare's data. Section \ref{sec:model} discusses the physical model used in calculating the SEDs and light curves
of the blazars and the method to extract the expected number of neutrinos detected by IceCube and KM3NeT for each flare from these simulations. In Section \ref{sec:results} the results of the modelling are discussed.

\section{Data Analysis} \label{sec:data}
\subsection{Swift XRT data}
The data from the Swift XRT were reduced,
calibrated, and cleaned using the XRT-PIPELINE in HEASOFT 6.15.1. Grades 0–12 were used for observations in photon counting mode and grade 0 for those in
window-timing mode \citep{Kreter2020}. Spectral fitting was performed in XSPEC, using a power-law spectrum with absorption modelled by the Tuebingen-Boulder ISM absorption model, \verb|tbabs| \citep{Wilms2000}. For this absorption, the photoelectric absorption cross-sections of \cite{Verner1996} and relative abundances in \cite{Wilms2000} were used. 

\subsection{Fermi-LAT Data}
All analysis of the Fermi-LAT data was performed using \verb|fermipy v1.2| \citep{Wood2018}. The instrument response function used was \verb|P8R2_SOURCE_V6|. The Galactic diffuse emission model \verb|gll_iem_v07.fits| was used in conjunction with the isotropic diffuse model \verb|iso_P8R3_SOURCE_V3_v1.txt|. \revOne{7}{As the variability of these flares was often less than 3 days the Fermi Lightcurve Repository proved unsuitable for use in this work as it does not have the required time resolution.}
All sources were fit according to their respective models in the 4FGL catalogue \citep{Abdollahi2020}. For each flare listed in \cite{Kreter2020} starting on t$_{\text{min}}$ [MJD] and ending on date t$_{\text{max}}$ [MJD] a light curve was generated between t$_{\text{min}}$-1~d and t$_{\text{max}}$+1~d, using either 6,12 or 24-hour binning depending on the length of the flare. An SED was then generated for the time bin within the range t$_{\text{min}}$ to t$_{\text{max}}$ with the highest photon flux between 100 MeV and 300 GeV.

\section{Model Description} \label{sec:model}
 The calorimetric approach used in \cite{Kreter2020} estimates the integrated neutrino energy flux as directly proportional to the integrated $\gamma$-ray flux. Using this method the maximum number of neutrinos a blazar flare can create is given by
\begin{equation}
    N_\nu^{\text{max}}=A_{\text{Eff}}\cdot\left(\frac{\phi_\gamma}{E_\nu}\right)\cdot\Delta t
    \label{eq:calorimetric_max}
\end{equation} 
Here $\phi_{\gamma}$ represents the integrated X-ray -- $\gamma$-ray flux between 10 keV and 20 GeV, $A_{\text{Eff}}$ the energy-dependent effective area of the IceCube detector, $E_\nu$ the energy of the neutrino, and $\Delta t$ the duration of the flare. A flux of mono-energetic neutrinos with energy $E_\nu=2$ PeV was used in \cite{Kreter2020} in order to be comparable to \cite{Kadler2016}. 
The maximum number of neutrino events is then scaled down by an empirical factor of 0.025 in order to provide a more realistic estimate of neutrino events.

In this work, the blazar flares were simulated using OneHaLe, a code for a time-dependent one-zone lepto-hadronic blazar model. This code assumes a homogeneous, spherical emission region with a time-dependent injection of protons and electrons in power-law distributions, subject to a disordered magnetic field. 
To simulate this emission region the code solves the Fokker-Planck equation,  
\begin{equation}
    \pdv{n_{i}(\gamma,t)}{t}=\pdv{}{\gamma}\left(\frac{\gamma^2}{(a+2)t_{\text{acc}}}\pdv{n_{i}(\gamma,t)
}{\gamma}\right)-\pdv{}{\gamma}\left(\dot{\gamma}\cdot n_{i}(\gamma,t)\right)+Q_{i}(\gamma,t)-\frac{n_{i}(\gamma,t)
}{t_{\text{esc}}}-\frac{n_{i}(\gamma,t)
}{\gamma t_{\text{decay}}}
\label{eq:FP_gamma}
\end{equation}
for electrons, protons, muons and charged pions.
The first term on the right-hand-side of Eq. \ref{eq:FP_gamma} accounts for second-order Fermi acceleration of the particles, while the second term accounts for first-order Fermi acceleration and radiative as well as adiabatic cooling. The injection of new particles is represented by the term $Q_{i}(\gamma, t)$. Terms four and five represent the escape and decay of particles, respectively. The term $\dot{\gamma}$ encompasses all energy losses resulting from radiative output and adiabatic cooling \cite[see, e.g.,][for details]{Boettcher2012,Cerruti2020, Zacharias2021, Zacharias2022}.

Using archival data from the SSDC SED builder (\url{https://tools.ssdc.asi.it/SED/}), parameters were found to model all sources in their quiescent states. To decrease the amount of free variables, values from literature were incorporated where possible, and scaling relationships were used for the radii and luminosities of the Broad Line Region (BLR) and Dusty Torus (DT).
Once an appropriate parameter set was found, this steady-state solution was then perturbed in order to reproduce the light curves observed by the Fermi-LAT. In conjunction with the light curves, the SEDs of the blazars at the point of maximum flux were also fitted. These perturbations are modelled using a modified version of the default OneHaLe perturbations. The perturbation is treated as an additional power-law proton spectrum injected alongside the existing proton injection spectrum. This additional power law has the form 
\begin{equation}
    Q_{p}'(\gamma,t)=Q_{0,p}'(t)\gamma^{-\alpha_p'}H(\gamma:\gamma_{min}',\gamma_{max}')
\end{equation}
with $Q_{0,p}'(t)$ evolving as a symmetric Gaussian and $H$ being the Heaviside function. Once the flare light curves and SEDs could be reproduced, the total number of observable neutrinos per time interval was calculated using the energy- and angle-dependent effective areas of IceCube \citep{Aartsen2013} as well as the \revOne{8}{proposed energy-dependent effective areas of the completed KM3NeT}  \citep{adrianmartinez2016} for the neutrino spectra of each flavour. This observable neutrino flux was integrated over the time intervals calculated using an asymmetric Gaussian and summed over flavours to calculate the total neutrino count for each flare.

The total number of neutrinos produced was then treated as the mean number of events in a Poisson distribution to calculate the probability of neutrinos from these flares being detected.
\begin{equation}
    P_{Det}=P(k\geq 1)=1-P(0)=1-e^{-N_{\nu}}
    \label{eq:poisson}
\end{equation}
where $k$ is the number of neutrinos detected and $N_{\nu}$ is the total number of neutrinos expected.

\subsection{Sources}
\revOne{1}{To compare results with \citep{Kreter2020}, the same sources were used in this work.}Parameter values that could be sourced from existing literature were used wherever possible to narrow down the number of free parameters, including the mass of the central black hole, the accretion disk luminosity and details regarding the relativistic motion of the jet. For sources where parameter values could not be found, values similar to the other sources in the sample were used. For the distances from the central black hole to the sources of external radiation fields (BLR and DT), scaling relationships from \cite{Ghisellini2009} were used. These relate the radii of both the BLR and the DT as proportional to the square-root of the luminosity of the accretion disk: 
\begin{equation}R_{\text{BLR}}=10^{17}\sqrt{L_{\text{AD},45}}
\end{equation}
\begin{equation}R_{\text{DT}}=2.5\times10^{18}\sqrt{L_{\text{AD},45}}
\end{equation} 
where $L_{\text{AD},45}$ is the luminosity of the accretion disk in units of $10^{45}$ erg/s. The BLR and DT were both set to a luminosity of 10\% of the accretion disk luminosity. The temperature of the DT was set at 2000~K as any temperature above this would cause the dust to sublimate \citep{Ghisellini2013}. The temperature of the BLR was kept constant at $10^4$~K following \cite{Ghisellini2013}. The radius of the emission region is constrained by 
\begin{equation}
    R_{\text{blob}}<\delta\frac{ct_{\text{var}}}{1+z}
\end{equation}
where $t_{\text{var}}$ is the flux-doubling timescale in the observer frame and the Doppler factor of the emission region is $\delta =  (\Gamma[1-\beta\cos\theta])^{-1}$. The size of the emission region can also be used to constrain the maximum energy of injected particles, as for a particle to be confined within the region, the Larmor radius of that particle must be smaller than the emission region size.

\subsubsection{PKS 1510-089 (32 Flares)}
PKS 1510-089 is an FSRQ located at a redshift of $z=0.36$ \citep{Thompson1990}, with a black-hole mass of $4.46\times 10^{8}$ M$_\odot$ \citep{Woo2002} and an accretion disk luminosity of $10^{46}$ erg/s \citep{Pucella2008}. The viewing angle and bulk Lorentz factor (BLF) of the jet have been sourced from \cite{Jorstad2017} as 1.2\textdegree and 22.5, respectively, giving a Doppler factor of $\delta=37$. This makes this the fastest jet in the sample. Within the dates included in this sample, PKS 1510-089 is observed to have a doubling time of approximately 0.6~days, implying that the spherical emission region has a radius smaller than $4.2\times 10 ^{16}$ cm.

\subsubsection{3C 279 (10 Flares)}
3C 279 is located at a redshift of $z= 0.53$ \citep{Marziani1996}. The mass of the central black hole is taken as $2.7\times10^8 \, M_\odot$ \citep{Woo2002} along with an accretion disk luminosity of $2\times 10^{45}$ erg/s \citep{Pian2005}. The viewing angle and BLF of the jet have been sourced from \cite{Jorstad2017} as 1.9\textdegree and 13.3, respectively, giving a Doppler factor of $\delta=22$. 
A flux doubling time of 0.12 days has been observed in the $\gamma$-ray light curve of 3C~279. This implies that the spherical emission region must have a radius smaller than 4.5$\times 10 ^{15}$~cm.

\subsubsection{PKS 1424-41 (3 Flares)}
PKS 1424-41 is located at redshift $z=1.52$ \citep{Murphy2019}. The viewing angle and BLF of the jet have been taken as roughly the mean of the other jets in the sample, i.e., 1.35\textdegree and 18, respectively. Within the flares analysed a doubling time of approximately 0.1 days is observed. This, in conjunction with the assumed Doppler factor, constrains the radius of the emission to less than 3.2$\times 10 ^{15}$ cm. The accretion disk luminosity is taken as $2\times10^{45}$ erg/s \citep{Pian1999}.

\subsubsection{PKS 1329-049 (2 Flares)}
PKS 1329-049, located at a redshift of $z=2.2$ \citep{Thompson1990}, exhibits a flux doubling time of approximately 0.4 days within the flares analysed. The jet viewing angle and BLF are assumed to be the same as those for PKS 1424-41 ($\theta_{\text{jet}}=1.35\text{\textdegree}$ and $\Gamma_{\text{jet}}=18$). This implies that the spherical emission region must have a radius smaller than 1.0$\times 10 ^{16}$ cm.

\subsubsection{PKS 0402-362 (1 Flare)}
PKS 0402-362 is located at a redshift of $z=1.42$ \citep{Jones2009}.
The viewing angle and BLF of the jet have been assumed in the same manner as for PKS 1329-049 and PKS 1424-41. PKS 0402-362 shows a flux doubling time of 0.2 days, implying a radius smaller than $6.5 \times 10^{15}$~cm.
The mass of the central black hole has been 
determined as $3.16\times10^9 \, M_\odot$ \citep{Decarli2011}.

\subsubsection{CTA 102 (1 Flare)}
CTA 102 is located at a redshift of $z=1.03$ \citep{Monroe2016} and has a central black hole mass of $8.51\times10^8 \, M_\odot$\citep{Zamaninasab2014}. It has a jet viewing angle and BLF of 1.6\textdegree and 21.4, respectively \citep{Jorstad2017}, which gives a Doppler factor of $\delta=32$. CTA 102 exhibits a flux doubling time of 0.11 days within the flare sampled here. This implies that the spherical emission region must have a radius smaller than $4.4 \times 10^{15}$~cm.

\subsubsection{3C 454.3 (1 Flare)}
3C 454.3 is one of the brightest $\gamma$-ray sources in the sky. It is located at a redshift of $z=0.86$ \citep{Koss2022} and has a black hole mass of $1.47 \times 10^9 \, M_\odot$ \citep{Woo2002}.
The viewing angle and BLF of the jet have been sourced from \cite{Jorstad2017} as 0.7\textdegree and 13.8, respectively, making it the most closely aligned jet in the sample. This yields a Doppler factor of $\delta=27$. The accretion disk luminosity is assumed to be $1.7\times10^{47}$ erg/s \citep{Raiteri2008}.
Within the flare studied here, 3C~454.3 shows a flux doubling time on the order of approximately 1 day. At other times, however, variability on the order of 2 hours has been observed in the optical band \citep{Weaver2019}. This implies that the spherical emission region must have a radius smaller than $3.1 \times 10^{15}$~cm.

\section{Results} \label{sec:results}

\begin{figure}[ht]
    \centering
    \includegraphics[width=0.85\columnwidth]{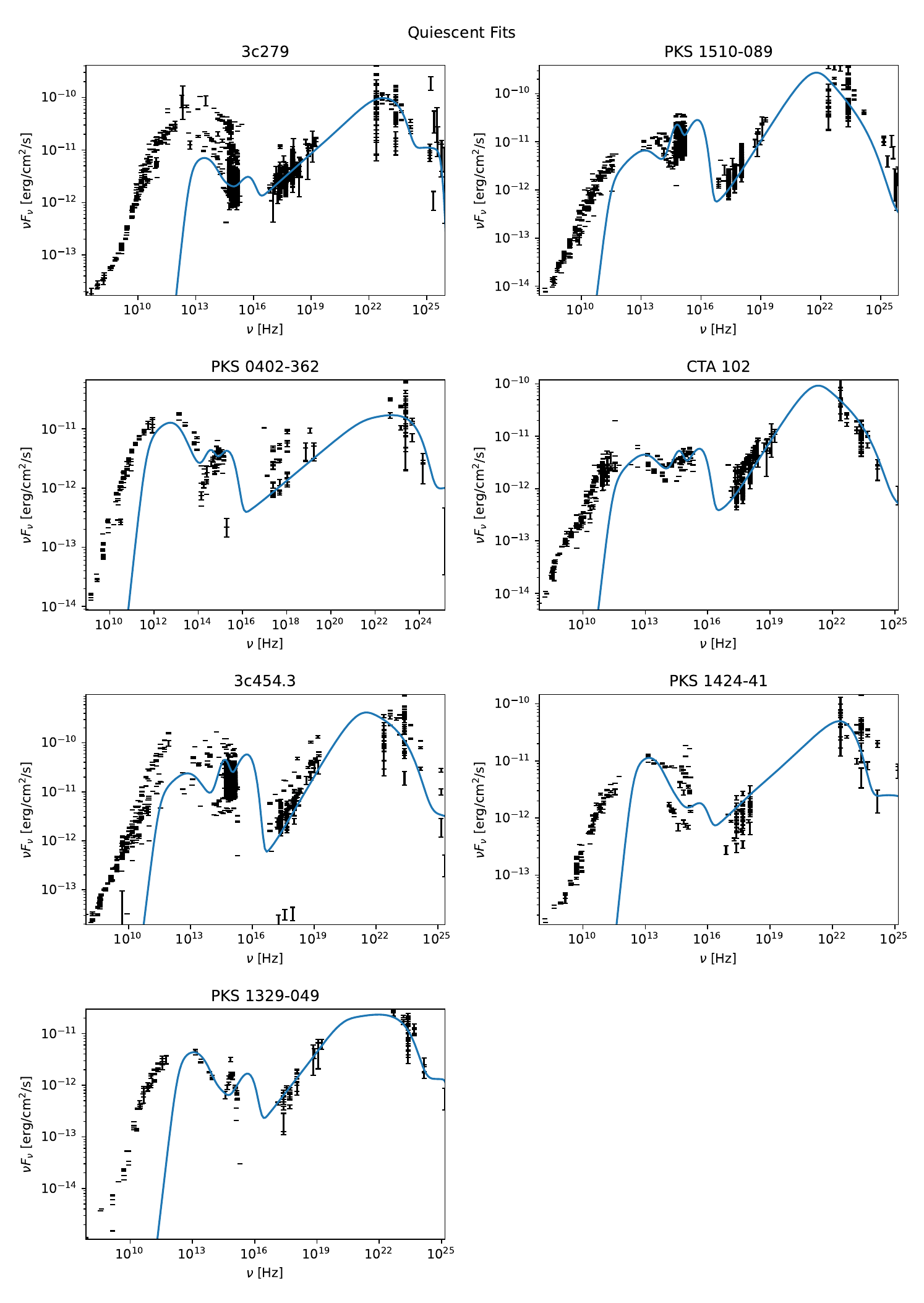}    \caption{Lepto-hadronic fits for each source in our sample in their quiescent state. See Table \ref{table:quiescent parameters} for parameters.}
    \label{fig:quiescent_fits}
\end{figure}

\begin{figure}[ht]
    \centering
    \includegraphics[width=1.0\columnwidth]{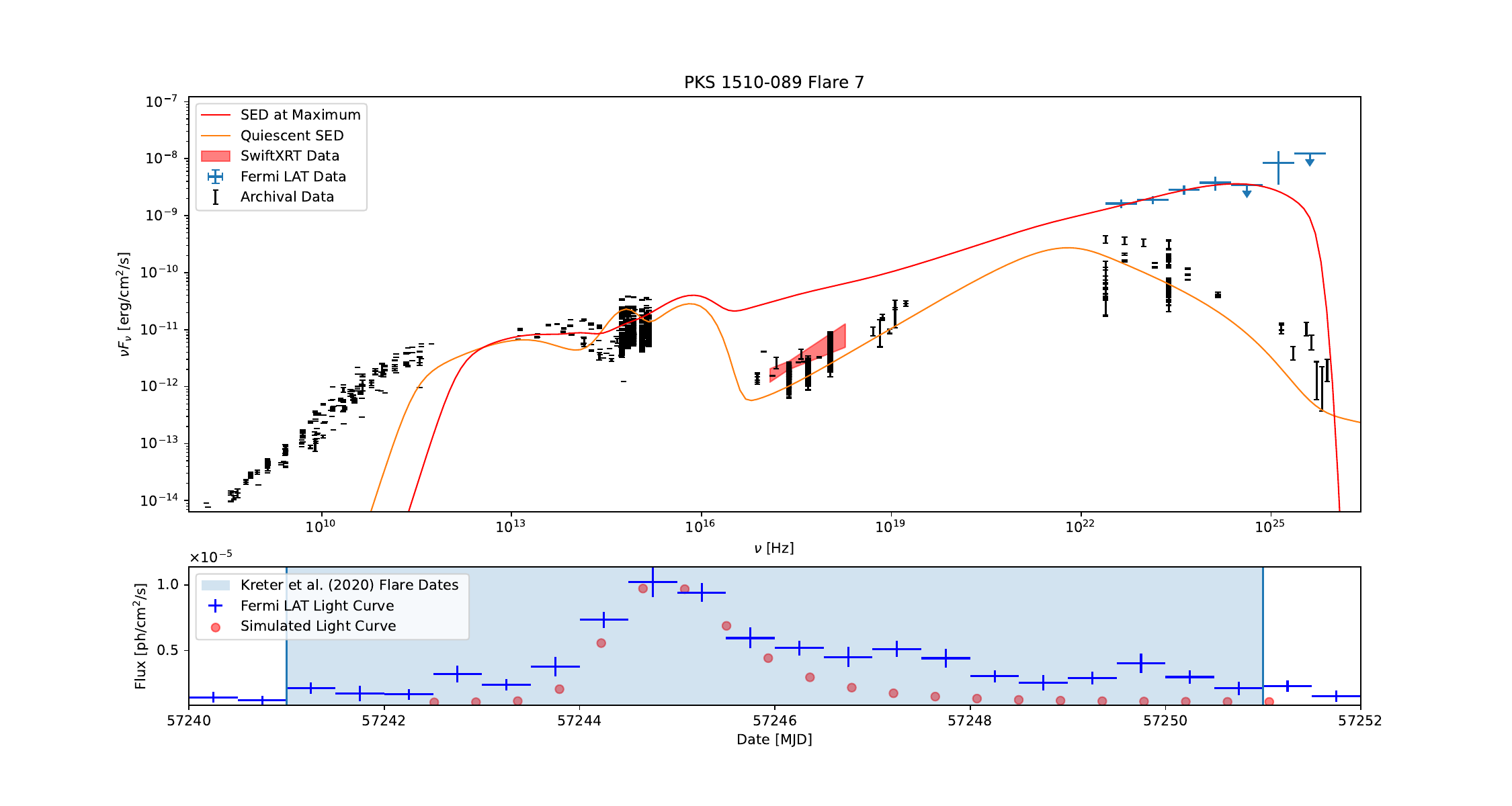}
    \caption{An example of a fit SED and light curve for a flaring event (Flare 7) of PKS 1510-089. In the top panel the quiescent SED fit is shown in orange and the SED fit at the peak of the flare is shown in red. The photon-flux light curve from Fermi-LAT is shown in the bottom panel in blue, along with the simulated light curve in red.}
    \label{fig:exmaple_SED_and_lightcurve}
\end{figure}

The quiescent-state fits found for each blazar in the \cite{Kreter2020} sample are shown in Fig. \ref{fig:quiescent_fits}, with their associated parameter sets listed in Table \ref{table:quiescent parameters}. For each flare the respective source's quiescent state was perturbed in order to reproduce the Fermi-LAT light curve as well as the SED at the point of maximal flux. An example of one of these fits is shown in Fig. \ref{fig:exmaple_SED_and_lightcurve}, with the original dates from \cite{Kreter2020} marked by the shaded region in the light curve plot. Table \ref{table:flare parameters} shows the parameters for each flare, as well as the total neutrino counts for both IceCube and KM3NeT. Flares 1, 22 and 35 were excluded from this due to difficulties in fitting: Attempts to model flares 1 and 22  both produced extremely high optical to X-ray fluxes, caused by runaway pair production. Once Flare 35 was analysed with sub-day binning, the Fermi-LAT light curve was found to be unusable for this type of study due to the lack of a discernible flaring event during the dates specified in \cite{Kreter2020} .

\begin{figure}[ht]
    \centering
    \includegraphics[width=\columnwidth]{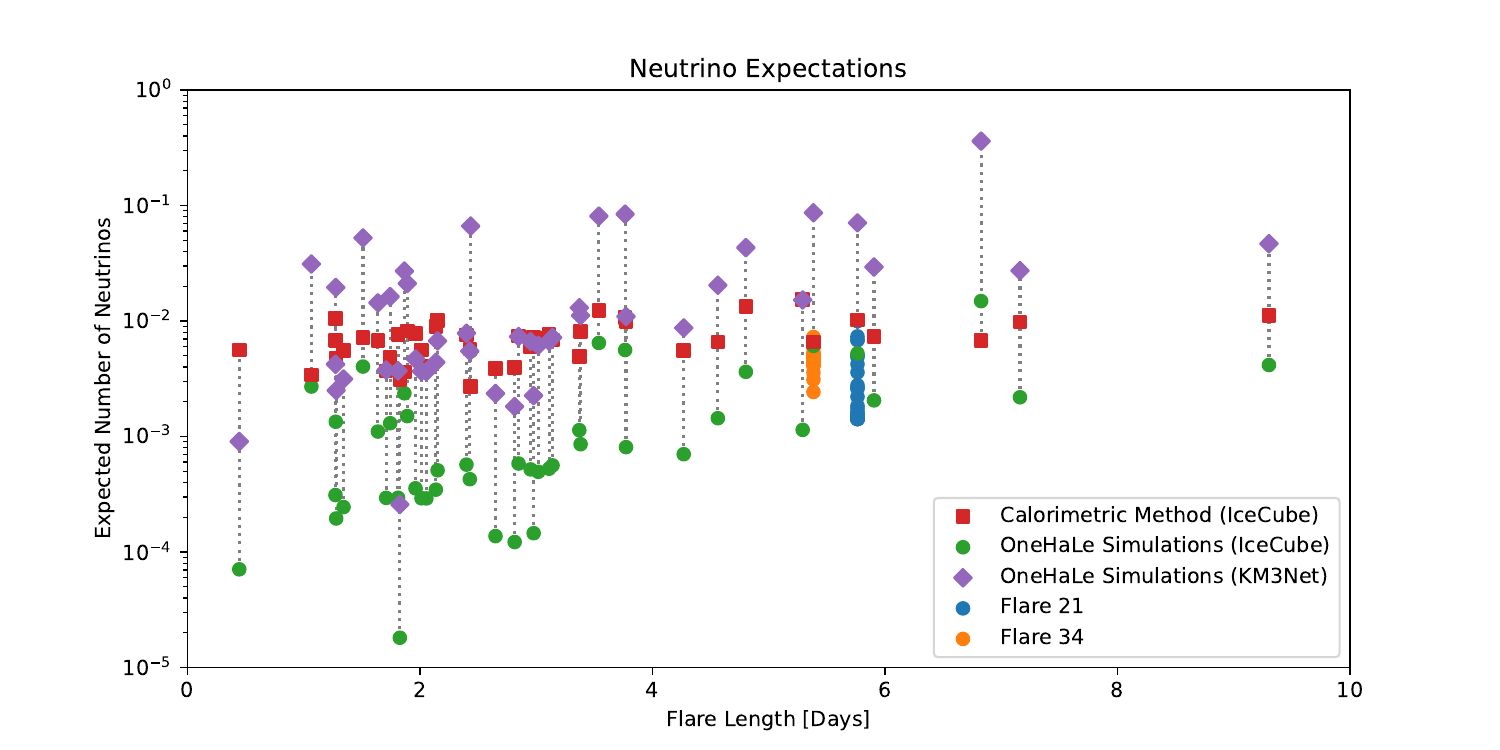}
    \caption{Neutrino count expectations for OneHaLe simulations of the sample of blazar flares. Expected counts for the IceCube Neutrino Observatory are shown in green, those for KM3NeT are shown in purple. The expected neutrino counts using the calorimetric method are shown in red. The orange and blue dots represent the results from the grid scans of flare 34 and flare 21 respectively.}
    \label{fig:neutrino_output_summary}
\end{figure}

The analysis showed that the expected IceCube neutrino counts, as shown in Fig. \ref{fig:neutrino_output_summary}, fall below the results from \citep{Kreter2020}, with most counts leading to a probability of detection of less than 
1~\%. However, when using the effective areas for KM3NeT, the probability of detection reaches nearly 10~\% in some cases.


\begin{table}[ht]
\begin{center}
\caption{Parameters for quiescent SED fits using OneHaLe}
\begin{tabular}{ l|c|ccccccc }

&Units&PKS&3C 279&PKS&CTA 102&3C 454.3&PKS &PKS\\
& &1510-089 & &0402-362& & &1424-41&1329-049\\
\hline
$z$&&0.4&0.5&1.4&1.0&0.9&1.5&2.1\\
$M_{bh}$&[$10^{8} M_{\odot}$]&4.5&2.7&31.6&8.5&14.7&4.0&4.0\\
$B$&[G]&70&150&70&80&150&100&100\\
$R_{blob}$&[cm]& $4.0\!\times\!10^{15}$& $1.0\!\times\!10^{15}$& $4.0\!\times\!10^{15}$& $3.0\!\times\!10^{15}$& $2.0\!\times\!10^{15}$& $5.0\!\times\!10^{15}$& $5.0\!\times\!10^{15}$\\
$\Gamma$&&22.5&13.3&18.0&21.4&13.8&18.0&18.0\\
$\theta_{jet}$&\textdegree& $2.1\!\times\!10^{-2}$& $3.3\!\times\!10^{-2}$& $2.4\!\times\!10^{-2}$& $2.8\!\times\!10^{-2}$& $1.2\!\times\!10^{-2}$& $2.4\!\times\!10^{-2}$& $2.4\!\times\!10^{-2}$\\
$\delta$&&36.8&22.3&30.5&31.5&26.8&30.5&30.5\\
$z_0$&[cm]& $1.7\!\times\!10^{18}$& $7.8\!\times\!10^{17}$& $3.5\!\times\!10^{18}$& $2.8\!\times\!10^{18}$& $7.2\!\times\!10^{18}$& $2.1\!\times\!10^{18}$& $3.7\!\times\!10^{18}$\\
$\frac{\eta R}{c}$&&5&6&5&4&5&3&3\\
$\frac{acc}{esc}$&&5.0&1.2&5.0&2.5&2.2&2.0&2.0\\
$p_e$&&2.9&3.5&3.8&2.9&3.3&3.5&3.8\\
$\gamma_{min,e}$&&42&50&47&43&34&50&50\\
$\gamma_{max,e}$&& $1.0\!\times\!10^{6}$& $1.0\!\times\!10^{6}$& $1.0\!\times\!10^{6}$& $1.0\!\times\!10^{6}$& $1.0\!\times\!10^{6}$& $1.0\!\times\!10^{9}$& $1.0\!\times\!10^{9}$\\
$L_e$&[$\frac{\text{erg}}{\text{s}}$]& $7.0\!\times\!10^{39}$& $1.3\!\times\!10^{41}$& $8.8\!\times\!10^{41}$& $1.0\!\times\!10^{41}$& $8.0\!\times\!10^{41}$& $8.2\!\times\!10^{41}$& $7.3\!\times\!10^{41}$\\
$p_p$&&1.7&2.2&2.4&1.7&1.5&2.3&1.7\\
$\gamma_{min,p}$&&100.1&10.1&100.1&1.1&1.1&1.1&1.1\\
$\gamma_{max,p}$&& $3.0\!\times\!10^{7}$& $6.0\!\times\!10^{7}$& $3.2\!\times\!10^{8}$& $3.0\!\times\!10^{7}$& $3.0\!\times\!10^{7}$& $1.0\!\times\!10^{8}$& $1.0\!\times\!10^{7}$\\
$L_p$&[$\frac{\text{erg}}{\text{s}}$]& $1.4\!\times\!10^{43}$& $7.5\!\times\!10^{44}$& $4.4\!\times\!10^{45}$& $3.0\!\times\!10^{44}$& $3.5\!\times\!10^{44}$& $5.9\!\times\!10^{45}$& $3.6\!\times\!10^{44}$\\
$\frac{L}{L_{Edd}}$&& $1.7\!\times\!10^{-1}$& $5.7\!\times\!10^{-2}$& $1.0\!\times\!10^{-1}$& $2.3\!\times\!10^{-1}$& $8.9\!\times\!10^{-1}$& $2.9\!\times\!10^{-1}$& $8.7\!\times\!10^{-1}$\\
$R_{blr}$&[cm]& $3.2\!\times\!10^{17}$& $1.4\!\times\!10^{17}$& $6.4\!\times\!10^{17}$& $5.0\!\times\!10^{17}$& $1.3\!\times\!10^{18}$& $3.9\!\times\!10^{17}$& $6.7\!\times\!10^{17}$\\
$L_{blr}$&[$\frac{\text{erg}}{\text{s}}$]& $1.0\!\times\!10^{45}$& $2.0\!\times\!10^{44}$& $4.1\!\times\!10^{45}$& $2.5\!\times\!10^{45}$& $1.7\!\times\!10^{46}$& $1.5\!\times\!10^{45}$& $4.6\!\times\!10^{45}$\\
$T_{blr}$&[K]&10000&10000&10000&10000&10000&10000&10000\\
$R_{DT}$&[cm]& $3.2\!\times\!10^{18}$& $1.4\!\times\!10^{18}$& $6.4\!\times\!10^{18}$& $5.0\!\times\!10^{18}$& $1.3\!\times\!10^{19}$& $3.9\!\times\!10^{18}$& $6.7\!\times\!10^{18}$\\
$L_{DT}$&[$\frac{\text{erg}}{\text{s}}$]& $1.0\!\times\!10^{45}$& $2.0\!\times\!10^{44}$& $4.1\!\times\!10^{45}$& $2.5\!\times\!10^{45}$& $1.7\!\times\!10^{46}$& $1.5\!\times\!10^{45}$& $4.6\!\times\!10^{45}$\\
$T_{DT}$&[K]&2000&2000&2000&2000&2000&2000&2000
\label{table:quiescent parameters}
\end{tabular}
\end{center}
\end{table}

\vbox{

\begin{longtable}{ c|cccccccccc}
\caption{Parameters for the additional proton power-law injections for the modelled flares. The last 4 columns list the expected numbers of neutrinos to be detected by IceCube and KM3NeT and the Poissonian probability of detection. }\\
Flare&Source&$\gamma_{min}$&$\gamma_{max}$&$\alpha$&$L_{p}$&$\Delta t$&$N_{\nu}^{IceCube}$&$P_{Det}^{IceCube}$&$N_{\nu}^{KM3NeT}$&$P_{Det}^{KM3NeT}$\\
& & & & & [$\frac{\text{erg}}{\text{s}}$]& [Days]&&\%&&\%\\
\hline
2&PKS 1510-089&$1.0\times10^{2}$&$2.5\times10^{8}$&1.7&$5.6\times10^{44}$&5.29&$1.1\times10^{-3}$&0.114&$1.5\times10^{-2}$&1.509\\
3&PKS 1510-089&$1.0\times10^{6}$&$1.5\times10^{8}$&2.5&$2.8\times10^{44}$&4.8&$3.6\times10^{-3}$&0.362&$4.3\times10^{-2}$&4.229\\
4&PKS 1510-089&$1.0\times10^{3}$&$1.0\times10^{8}$&1.7&$1.2\times10^{44}$&2.15&$5.1\times10^{-4}$&0.051&$6.7\times10^{-3}$&0.669\\
5&3C 279&$1.0\times10^{1}$&$1.0\times10^{8}$&1.2&$2.8\times10^{44}$&1.28&$1.3\times10^{-3}$&0.134&$2.0\times10^{-2}$&1.933\\
6&3C 279&$3.0\times10^{7}$&$4.0\times10^{7}$&1.2&$3.1\times10^{44}$&3.76&$5.6\times10^{-3}$&0.559&$8.4\times10^{-2}$&8.071\\
7&PKS 1510-089&$1.0\times10^{2}$&$3.0\times10^{9}$&2.3&$1.8\times10^{45}$&7.16&$2.2\times10^{-3}$&0.218&$2.7\times10^{-2}$&2.695\\
8&3C 279&$1.0\times10^{1}$&$1.0\times10^{7}$&1.3&$9.6\times10^{44}$&3.54&$6.5\times10^{-3}$&0.644&$8.1\times10^{-2}$&7.771\\
9&PKS 1510-089&$1.0\times10^{7}$&$1.0\times10^{8}$&2.7&$3.6\times10^{43}$&3.77&$8.1\times10^{-4}$&0.081&$1.1\times10^{-2}$&1.087\\
10&PKS 1510-089&$1.0\times10^{7}$&$2.0\times10^{8}$&2.7&$4.0\times10^{43}$&2.4&$5.7\times10^{-4}$&0.057&$7.8\times10^{-3}$&0.781\\
11&PKS 1510-089&$1.0\times10^{7}$&$1.0\times10^{8}$&2.9&$3.1\times10^{43}$&1.27&$3.1\times10^{-4}$&0.031&$4.2\times10^{-3}$&0.421\\
12&PKS 1510-089&$1.0\times10^{2}$&$6.0\times10^{7}$&1.7&$1.7\times10^{44}$&3.38&$8.6\times10^{-4}$&0.086&$1.1\times10^{-2}$&1.109\\
13&PKS 1510-089&$1.0\times10^{2}$&$2.0\times10^{8}$&2.0&$5.1\times10^{43}$&1.81&$2.9\times10^{-4}$&0.029&$3.7\times10^{-3}$&0.37\\
14&PKS 1510-089&$1.0\times10^{7}$&$1.0\times10^{9}$&2.8&$4.1\times10^{43}$&1.96&$3.6\times10^{-4}$&0.036&$4.7\times10^{-3}$&0.468\\
15&PKS 1510-089&$1.0\times10^{2}$&$1.5\times10^{8}$&2.2&$4.1\times10^{44}$&3.02&$4.9\times10^{-4}$&0.049&$6.2\times10^{-3}$&0.614\\
16&PKS 1510-089&$1.0\times10^{5}$&$2.0\times10^{9}$&2.3&$2.4\times10^{44}$&4.27&$7.0\times10^{-4}$&0.07&$8.7\times10^{-3}$&0.868\\
17&PKS 1510-089&$1.0\times10^{2}$&$2.0\times10^{8}$&1.7&$6.0\times10^{42}$&2.01&$2.9\times10^{-4}$&0.029&$3.7\times10^{-3}$&0.366\\
18&PKS 1510-089&$1.0\times10^{2}$&$8.0\times10^{7}$&1.7&$5.2\times10^{43}$&2.95&$5.2\times10^{-4}$&0.052&$6.6\times10^{-3}$&0.662\\
19&PKS 1510-089&$1.0\times10^{4}$&$5.0\times10^{7}$&2.0&$5.1\times10^{43}$&27.44&$6.4\times10^{-3}$&0.638&$8.0\times10^{-2}$&7.658\\
20&PKS 1510-089&$1.0\times10^{6}$&$1.0\times10^{7}$&2.0&$1.3\times10^{44}$&9.3&$4.2\times10^{-3}$&0.415&$4.7\times10^{-2}$&4.564\\
21&PKS 1510-089&$1.0\times10^{2}$&$4.0\times10^{8}$&2.5&$1.8\times10^{46}$&5.76&$5.2\times10^{-3}$&0.522&$7.1\times10^{-2}$&6.828\\
23&PKS 1510-089&$1.0\times10^{2}$&$3.0\times10^{8}$&1.7&$9.3\times10^{42}$&0.45&$7.1\times10^{-5}$&0.007&$9.1\times10^{-4}$&0.091\\
24&PKS 1510-089&$1.0\times10^{2}$&$7.0\times10^{7}$&1.7&$3.1\times10^{43}$&3.11&$5.3\times10^{-4}$&0.053&$6.7\times10^{-3}$&0.666\\
25&PKS 1510-089&$1.0\times10^{2}$&$1.0\times10^{8}$&1.7&$4.7\times10^{43}$&2.43&$4.3\times10^{-4}$&0.043&$5.5\times10^{-3}$&0.547\\
26&PKS 1510-089&$1.0\times10^{6}$&$1.0\times10^{8}$&1.1&$1.2\times10^{43}$&1.28&$2.0\times10^{-4}$&0.02&$2.5\times10^{-3}$&0.25\\
27&3C 279&$1.0\times10^{1}$&$6.0\times10^{7}$&2.0&$1.1\times10^{45}$&1.89&$1.5\times10^{-3}$&0.15&$2.1\times10^{-2}$&2.089\\
28&PKS 0402-362&$3.2\times10^{7}$&$3.2\times10^{8}$&3.0&$2.0\times10^{45}$&2.98&$1.5\times10^{-4}$&0.015&$2.3\times10^{-3}$&0.226\\
29&PKS 1510-089&$1.0\times10^{2}$&$8.0\times10^{7}$&1.7&$1.9\times10^{43}$&2.14&$3.5\times10^{-4}$&0.035&$4.4\times10^{-3}$&0.438\\
30&PKS 1510-089&$1.0\times10^{2}$&$1.0\times10^{8}$&1.7&$2.0\times10^{43}$&1.71&$2.9\times10^{-4}$&0.029&$3.8\times10^{-3}$&0.375\\
31&CTA 102&1.1&$4.0\times10^{8}$&2.5&$1.5\times10^{24}$&2.44&$5.5\times10^{-3}$&0.545&$6.6\times10^{-2}$&6.421\\
32&3C 279&$1.0\times10^{6}$&$7.5\times10^{7}$&2.2&$3.1\times10^{44}$&4.56&$1.4\times10^{-3}$&0.144&$2.0\times10^{-2}$&2.018\\
33&PKS 1510-089&$1.0\times10^{2}$&$1.0\times10^{8}$&1.7&$2.0\times10^{43}$&1.34&$2.4\times10^{-4}$&0.024&$3.2\times10^{-3}$&0.315\\
34&3C 279&$1.0\times10^{1}$&$6.0\times10^{8}$&2.0&$7.1\times10^{44}$&5.38&$6.1\times10^{-3}$&0.607&$8.6\times10^{-2}$&8.281\\
36&3C 279&$1.0\times10^{7}$&$1.0\times10^{8}$&1.8&$1.6\times10^{44}$&5.91&$2.1\times10^{-3}$&0.205&$2.9\times10^{-2}$&2.895\\
37&PKS 1510-089&$1.0\times10^{2}$&$9.0\times10^{7}$&2.4&$2.2\times10^{46}$&1.64&$1.1\times10^{-3}$&0.11&$1.4\times10^{-2}$&1.426\\
38&3C 454.3&1.1&$1.0\times10^{8}$&1.5&$6.1\times10^{44}$&127.05&$1.4\times10^{-2}$&1.4&$1.9\times10^{-1}$&17.124\\
39&PKS 1510-089&$1.0\times10^{1}$&$1.0\times10^{8}$&2.4&$2.2\times10^{46}$&1.74&$1.3\times10^{-3}$&0.13&$1.6\times10^{-2}$&1.612\\
40&PKS 1510-089&$1.0\times10^{2}$&$8.0\times10^{7}$&2.0&$1.2\times10^{44}$&2.85&$5.8\times10^{-4}$&0.058&$7.3\times10^{-3}$&0.731\\
41&PKS 1424-41&1.1&$6.0\times10^{8}$&2.4&$5.3\times10^{47}$&2.65&$1.4\times10^{-4}$&0.014&$2.4\times10^{-3}$&0.236\\
42&3C 279&$1.0\times10^{1}$&$1.0\times10^{7}$&2.0&$2.9\times10^{45}$&1.51&$4.0\times10^{-3}$&0.402&$5.2\times10^{-2}$&5.111\\
43&PKS 1424-41&1.1&$6.0\times10^{8}$&2.3&$9.8\times10^{46}$&2.81&$1.2\times10^{-4}$&0.012&$1.8\times10^{-3}$&0.182\\
44&PKS 1510-089&$1.0\times10^{2}$&$2.0\times10^{8}$&2.8&$5.4\times10^{47}$&6.83&$1.5\times10^{-2}$&1.476&$3.6\times10^{-1}$&30.3\\
45&PKS 1510-089&$1.0\times10^{2}$&$8.0\times10^{7}$&1.7&$1.9\times10^{43}$&3.14&$5.6\times10^{-4}$&0.056&$7.2\times10^{-3}$&0.718\\
46&PKS 1510-089&$1.0\times10^{2}$&$3.0\times10^{8}$&1.7&$2.9\times10^{42}$&2.06&$2.9\times10^{-4}$&0.029&$3.6\times10^{-3}$&0.364\\
47&PKS 1329-049&1.1&$1.0\times10^{7}$&1.0&$1.0\times10^{26}$&1.07&$2.7\times10^{-3}$&0.27&$3.1\times10^{-2}$&3.066\\
48&PKS 1329-049&1.1&$1.0\times10^{7}$&1.5&$2.1\times10^{26}$&1.87&$2.4\times10^{-3}$&0.236&$2.7\times10^{-2}$&2.669\\
49&PKS 1424-41&1.1&$9.0\times10^{8}$&1.5&$5.4\times10^{44}$&1.83&$1.8\times10^{-5}$&0.002&$2.6\times10^{-4}$&0.026\\
50&PKS 1510-089&$1.0\times10^{2}$&$1.0\times10^{7}$&2.0&$4.6\times10^{44}$&3.37&$1.1\times10^{-3}$&0.113&$1.3\times10^{-2}$&1.294
\label{table:flare parameters}
\end{longtable}
}

As all fitting of the perturbation parameters was done by eye and there are substantial parameter degeneracies, the chosen set of parameters for each flare is not unique and may not be the one that maximizes the expected neutrino detection rate. 
In order to investigate the impact of these parameter degeneracies on the predicted neutrino detection rates, a grid scan was performed on two of the flares, namely Flare 21 and Flare 34. \revOne{2}{ Grid scans were used as exploring the parameter spaces fully was prohibitively expensive in terms of computational times} 
The results of both grid scans are included in Fig. \ref{fig:neutrino_output_summary}, with the results of all acceptable simulations presented in blue and orange. This revealed that the maximum and minimum neutrino counts do not differ by more than roughly one order of magnitude. This implies that, while each set of parameters may not necessarily represent the most optimistic neutrino prediction, the results presented do give an accurate indication of neutrino expectations. 
Flare 38, a month-long outburst from 3C 454.3, stands out within the sample as not only being the longest flare in the sample, but also producing the most neutrinos as a result of this length. The fitting in this work was only able to recreate the long term behaviour of this flare, and as such the true level of neutrino production may be even higher if the short term behaviour is included.

\section{Summary and Discussion}
In this work, we have modelled the flares of seven different blazars. Using a time-dependent lepto-hadronic code the steady-state quiescent solutions were perturbed to recreate the flare light cures as well as the SEDs during the flares. The neutrino flux produced during each of these flares was then convolved with the effective areas of the IceCube Neutrino Observatory and KM3NeT to calculate the expected neutrino counts for each detector during the flares. From the total neutrino counts, the probability of a positive detection (the probability of observing one or more neutrinos) was then calculated by treating the counts calculated above as the mean event rate in a Poisson distribution.
This analysis showed that, of the flares modelled, IceCube was unlikely to have made a definitive detection of any of the blazars as a point source. As many of the flares had a detection probability of less than 1~\% the non-detection of these blazars, even while in a flaring state, is the most probable result. These results therefore support the results and conclusions of \cite{Kreter2020} as for most flares the neutrino counts are overestimated by the calorimetric method by only a small factor.
For the KM3NeT detector, however, the results are more promising, with the detection probability of many flares being on the order of $\sim$10~\% . This implies that with multiple flares each year the probability of detecting a neutrino from these sources is high over extended periods of time.

\subsection{Limitations}
The flares here are sourced from \cite{Kreter2020}, which used daily binning for light curves, therefore excluding sub-day length flares. 
Furthermore, the fits presented in this work represent only one of a number of degenerate parameter choices and do not necessarily maximise the neutrino detection predictions. Separate fits may be found, even within a different parameter regime (for example, one with the high energy emission being produced by inverse-Compton radiation with only a sub-dominant hadronic component). The analysis performed using grid scans has, however, shown that if the high-energy emission is hadronically dominated, our predictions are expected to be accurate within an order of magnitude. 

The flares are also being modelled with very simple perturbations from the quiescent state, whereas in reality, flares may be a result of much more complex processes, possibly leading to different neutrino predictions. 
Some of our flare fit attempts produced excessive emission in the optical -- UV bands due to p$\gamma$ induced pair cascades, of magnitudes which are not seen in archival data for these sources. This shows that while these simple models for the flaring states can reproduce the spectral and temporal changes seen in $\gamma$-rays, future simultaneous optical -- UV observations are required to verify or rule out a hadronically dominated model for blazar flares.


\section{Conclusions}
This work has shown that even in a flaring state the sampled blazars do not produce neutrinos in high enough quantities to lead to definitive detection using current generation neutrino observatories, in particular IceCube. This is consistent with other work such as \cite{Yoshida2023} and indicates that while the presence of neutrinos from blazar jets cannot be ruled out, it is unlikely that they are predominantly the source of high-energy cosmic neutrinos\revOne{3}{, though they may still contribute a notable fraction.} The projections for KM3NeT neutrino counts show promise that future detectors may indeed be able to detect more individual blazars as neutrino point sources. 

 Our work shows that the calorimetric method, while overestimating the amount of neutrinos produced by a factor of $\sim 10$, still holds for the general trend of flaring proton-synchrotron blazars. If the empirical factor is adjusted accordingly then this method presents a much faster way of estimating neutrino expectations from flaring blazars than physical modelling.
\bibliography{bibliography2}{}
\bibliographystyle{aasjournal}
 


\end{document}